\documentclass[onecolumn,preprint,noshowkeys,noshowpacs]{revtex4}%
\usepackage{amssymb}
\usepackage{amsmath}
\usepackage{amsfonts}
\usepackage{graphicx}%
\providecommand{\U}[1]{\protect\rule{.1in}{.1in}}

\begin{document}
\title{Entanglement and quantum phase transition in the one-dimensional anisotropic
XY model }
\author{Fu-Wu Ma, Sheng-Xin Liu}
\author{Xiang-Mu Kong}
\thanks{Corresponding author}
\email{kongxm@mail.qfnu.edu.cn (X.-M. Kong)}
\affiliation{Shandong Provincial Key Laboratory of Laser Polarization and Information
Technology, Department of physics, Qufu Normal university, Qufu 273165, China}
\date{\today}

\begin{abstract}
In this paper the entanglement and quantum phase transition of the anisotropic
$s=1/2$ XY model are studied by using the quantum renormalization group
method. By solving the renormalization equations, we get the trivial fixed
point and the untrivial fixed point which correspond to the phase of the
system and the critical point, respectively. Then the concurrence between two
blocks are calculated and it is found that when the number of the iterations
of the renormalziation trends infinity, the concurrence develops two
staturated values which are associated with two different phases, i.e.,
Ising-like and spin-fluid phases. We also investigate the first derivative of
the concurrence, and find that there exists non-analytic behaviors at the
quantum critical point, which directly associate with the divergence of the
correlation length. Further insight, the scaling behaviors of the system are
analyzed, it is shown that how the maximum value of the first derivative of
the concurrence reaches the infinity and how the critical point is touched as
the size of the system becomes large.

\end{abstract}
\maketitle

\section{INTRODUCTION}

The quantum entanglement, as one of the most intriguing features of quantum
theory, has attracted much attention because its non-classical correlation can
be regarded as an essential resource in quantum communication and information
processing \cite{ma2000}. In view of the connection between the quantum
entanglement and quantum correlation \cite{J.S.1964}, the entanglement in some
many-body systems has been widely investigated
\cite{manybody1,manybody2,manybody3}. The entanglement, which is the main
difference between quantum and classical systems, also plays an important role
in the quantum phase transition (QPT). The QPT is induced by the change of an
external parameter or coupling constant \cite{S.Sachdev quantum}. This change
occurs at absolute zero temperature where the quantum fluctuations paly the
dominant role and all the thermal fluctuations get frozen.

The entanglement between two nearest-neighbor spins in one-dimensional XY
system with the transverse magnetic field was studied in Ref.
\cite{A.Osterloh2002}. The relation between the QPT and entanglement are
studied in this paper and it is found that the system exists QPT and the
scaling behavior in the vicinity of the critical point. After that, a great
deal of efforts have been devoted to investigating the entanglement and QPT in
some spin systems \cite{A.Wu.2004,G.Vidal2003,J.Vidal.2004,T.J.2002}. The XY
model is exactly solved by the Jordan-Wigner transform in Ref. \cite{T.J.2002}%
, in which the Ising model, as a special case of the XY model, exhibits QPT
and a maximum value for the next-nearest-neighbor entanglement at the critical
point. The entanglement properties and the spin squeezing of the ground state
of mutually interacting spins 1/2 in a transverse magnetic field are analyzed
and the system shows a cusplike singularity at the critical point in the
thermodynamical limit \cite{J.Vidal.2004}. To further discuss the QPT and
entanglement, the renormalization group (RG) method is introduced, such as
real space RG, Monte Carlo RG, and density-matrix RG \cite{rg1,rg2,rg3,rg4}.
The pairwise entanglement of the system is also discussed by means of quantum
renormalization group (QRG) method \cite{qrg1,qrg2}. Very recently, the
spin$-1/2$ Ising and Heisenberg models are studied by using the same method by
a group of Iran and found that the systems exist QPT \cite{A1,A2,A3,A4}. It is
also shown that the nonanalytic behavior of the entanglement and the scaling
behaviors closing to the quantum critical point are obtained.

The XY model is firstly discussed in Ref. \cite{E.Lieb1970}, and then has been
widely investigated recently \cite{xy1,xy2,xy3}. In this paper, we study the
quantum entanglement and QPT in the spin$-1/2$ XY model by the method of QRG.
It is found that the system exists QPT between the spin-fluid and Ising-like
phases. Further insight, the nonanalytic behavior of the entanglement and the
the scaling behavior of the system are gotten. The organization of this paper
is as follows. In Sec. \ref{dier} we briefly introduce the RG method and
obtain the fixed points of the XY model. The block-block entanglement is
obtained in Sec. \ref{disan} and in Sec. \ref{disi} we discuss the
non-analytic and the scaling behaviors of the entanglement. Sec. \ref{diwu} is
devoted to the conclusions.

\section{RENORMALIZATION OF THE XY MODEL \label{dier}}

Firstly, the QRG is introduced. Eliminating the degrees of freedom of the
system followed by an iteration is the main idea of QRG. The aim of the
iteration is that reduces the number of variables step by step until a more
manageable situation is reached. For this idea, the kadanoff's block approach
is implemented in this paper, because it is not only well suited to perform
analytical calculations but also easy to be extended to the higher dimensions
\cite{MA1996,High}. We consider three sites as a block, the kadanoff's block
approach is given in Fig. 1. In this way, the effective Hamiltonian $\left(
H^{eff}\right)  $ can be gotten which has structural similarities with the
original Hamiltonian $\left(  H\right)  $. The original Hilbert space is also
replaced by a reduced Hilbert space which acts on the renormalized subspace
\cite{Heff,MA1996}.

The Hamiltonian of the XY model on a periodic chain with N sites can be
written as%
\begin{equation}
H\left(  J,\gamma\right)  =\frac{J}{4}\sum_{i}^{N}\left[  \left(
1+\gamma\right)  \sigma_{i}^{x}\sigma_{i+1}^{x}+\left(  1-\gamma\right)
\sigma_{i}^{y}\sigma_{i+1}^{y}\right]  , \label{H}%
\end{equation}
where $J$ is the exchange coupling constant, $\gamma$ is the anisotropy
parameter, and $\sigma^{\alpha}$ $\left(  \alpha=x,y\right)  $ are Pauli
matrices. The XY model can be encompassed another two well-know spin models,
i.e., the Ising model for $\gamma=1$ and the XX model for $\gamma=0$. For
$0<\gamma\leq1$, it belongs to the Ising universality class \cite{E.Lieb1970}.

By using the Kadanoff's block approach, Eq. $\left(  \text{\ref{H}}\right)  $
can be written as
\begin{equation}
H=H^{A}+H^{AA}, \label{s}%
\end{equation}
where $H^{A}$ is the block Hamiltonian, $H^{AA}$ is the interblock
Hamiltonian. The specific forms of $H^{A}$ and $H^{AA}$\ are%
\begin{equation}
H^{A}=\sum_{L}^{N/3}h_{L}^{A}, \label{000}%
\end{equation}%
\begin{equation}
H^{AA}=\frac{J}{4}\sum_{L}^{N/3}\left[  \left(  1+\gamma\right)  \sigma
_{L,3}^{x}\sigma_{L+1,1}^{x}+\left(  1-\gamma\right)  \sigma_{L,3}^{y}%
\sigma_{L+1,1}^{y}\right]  , \label{11}%
\end{equation}
where%
\begin{equation}
h_{L}^{A}=\frac{J}{4}\left[  \left(  1+\gamma\right)  \left(  \sigma_{L,1}%
^{x}\sigma_{L,2}^{x}+\sigma_{L,2}^{x}\sigma_{L,3}^{x}\right)  +\left(
1-\gamma\right)  \left(  \sigma_{L,1}^{y}\sigma_{L,2}^{y}+\sigma_{L,2}%
^{y}\sigma_{L,3}^{y}\right)  \right]  , \label{5}%
\end{equation}
which is the $L$th block Hamiltonian.

In the terms of matrix product states \cite{Matrix}, the $L$th block
Hamiltonian can be exactly diagonalized and solved. We can obtain four
distinct eigenvalues which are doubly-degeneracy. Defining $\left\vert
\uparrow\right\rangle \ $and $\left\vert \downarrow\right\rangle $ as the
eigenstates of $\sigma^{z}$, both degenerate ground states are given as
follows:%
\begin{equation}
\left\vert \Phi_{0}\right\rangle =\frac{1}{2\sqrt{1+\gamma^{2}}}\left(
-\sqrt{1+\gamma^{2}}\left\vert \uparrow\uparrow\downarrow\right\rangle
+\sqrt{2}\left\vert \uparrow\downarrow\uparrow\right\rangle -\sqrt
{1+\gamma^{2}}\left\vert \downarrow\uparrow\uparrow\right\rangle +\sqrt
{2}\gamma\left\vert \downarrow\downarrow\downarrow\right\rangle \right)  ,
\label{6}%
\end{equation}%
\begin{equation}
\left\vert \Phi_{0}^{^{\prime}}\right\rangle =\frac{1}{2\sqrt{1+\gamma^{2}}%
}\left(  -\sqrt{2}\gamma\left\vert \uparrow\uparrow\uparrow\right\rangle
+\sqrt{1+\gamma^{2}}\left\vert \uparrow\downarrow\downarrow\right\rangle
-\sqrt{2}\left\vert \downarrow\uparrow\downarrow\right\rangle +\sqrt
{1+\gamma^{2}}\left\vert \downarrow\downarrow\uparrow\right\rangle \right)  .
\label{7}%
\end{equation}
The energy corresponding the ground states is%
\begin{equation}
E_{0}=-\frac{J}{\sqrt{2}}\sqrt{1+\gamma^{2}}. \label{JJ1}%
\end{equation}

For eliminating the higher energy of the system and retaining the lower, the
projection operator $T_{0}$ is composed by the lowest energy eigenstates of
the system. The relation between the original Hamiltonian and the effective
Hamiltonian can be given by the projection operator \cite{hh}, i.e.,
$H^{eff}=T_{0}^{+}HT_{0}$, where $T_{0}^{+}$\ is the hermitian operator of
$T_{0}$. In the effective Hamiltonian, we only consider the first order
correction in the perturbation theory. The effective Hamiltonian is%
\begin{equation}
H^{eff}=H_{0}^{eff}+H_{1}^{eff}=T_{0}^{+}H^{A}T_{0}+T_{0}^{+}H^{AA}T_{0}.
\label{hh}%
\end{equation}
The projection operator $T_{0}$ can be searched in a factorized form%
\begin{equation}
T_{0}=%
{\textstyle\prod_{i=1}^{N/3}}
T_{0}^{L}, \label{8}%
\end{equation}
where $T_{0}^{L}$ is the $L$th block, which is defined as%
\begin{equation}
T_{0}^{L}=\left\vert \Uparrow\right\rangle _{L}\left\langle \Phi
_{0}\right\vert +\left\vert \Downarrow\right\rangle _{L}\left\langle \Phi
_{0}\right\vert . \label{T}%
\end{equation}
The $\left\vert \Uparrow\right\rangle _{L}$ and $\left\vert \Downarrow
\right\rangle _{L}$ in the Eq. $\left(  \text{\ref{T}}\right)  $ are the
renamed states of $L$th block which can be seen as a new spin$-1/2$. The
renormalization of Pauli matrices are given by%
\begin{equation}
T_{0}^{L}\sigma_{i,L}^{\alpha}T_{0}^{L}=\eta_{i}^{\alpha}\sigma_{L}^{^{\prime
}\alpha}\ \ \left(  i=1,2,3;\ \ \alpha=x,y\right)  , \label{F}%
\end{equation}
where%
\begin{align}
\eta_{1}^{x}  &  =\eta_{3}^{x}=\frac{1+\gamma}{\sqrt{2\left(  1+\gamma
^{2}\right)  }},\ \ \eta_{2}^{x}=-\frac{\left(  1+\gamma\right)  ^{2}%
}{2\left(  1+\gamma^{2}\right)  },\nonumber\\
\eta_{1}^{y}  &  =\eta_{3}^{y}=\frac{1-\gamma}{\sqrt{2\left(  1+\gamma
^{2}\right)  }},\ \ \eta_{2}^{y}=-\frac{\left(  1-\gamma\right)  ^{2}%
}{2\left(  1+\gamma^{2}\right)  }. \label{JJ}%
\end{align}
Then, the effective Hamiltonian of the renormalized chain can be gotten,%
\begin{equation}
H^{eff}=\frac{J^{^{\prime}}}{4}\sum_{L}^{N/3}\left[  \left(  1+\gamma
^{^{\prime}}\right)  \sigma_{L}^{x}\sigma_{L+1}^{x}+\left(  1-\gamma
^{^{\prime}}\right)  \sigma_{L}^{y}\sigma_{L+1}^{y}\right]  , \label{H2}%
\end{equation}
where%
\begin{equation}
J^{^{\prime}}=J\frac{3\gamma^{2}+1}{2\left(  1+\gamma^{2}\right)  }%
,\ \ \gamma^{^{\prime}}=\frac{\gamma^{3}+3\gamma}{3\gamma^{2}+1}. \label{J}%
\end{equation}

In the above equations, not only do we get the nontrivial fixed point
$\gamma=0$, but also obtain the trivial fixed point $\gamma=1$ which specifies
the critical point of the system by solving $\gamma\equiv\gamma^{^{\prime}}$.
When $\gamma\rightarrow0$, the model falls into the universality class of XX
model corresponding to a spin-fluid phase; while for $\gamma\rightarrow1,$ the
system is in Ising-like phase. From above analyzing, it is found that there is
a phase boundary that separates the spin-fluid phase $\gamma=0$, from the
Ising-like phase $0<\gamma\leq1$.

\section{ENTANGLEMENT ANALYSIS \label{disan}}

We investigate the ground-state entanglement between two blocks of the XY
chain by the concept of concurrence \cite{C1,C2} and demonstrate how the
concurrence varies as the size of the block becomes large. We consider one of
the degenerate ground states to define the pure state density matrix. Thus,
the pure state density matrix is defined by%
\begin{equation}
\rho=\left\vert \Phi_{0}\right\rangle \left\langle \Phi_{0}\right\vert ,
\label{ppp}%
\end{equation}
where $\left\vert \Phi_{0}\right\rangle $ has been introduced in Eq.
(\ref{6}). The results is same if we consider the Eq. $\left(  \text{\ref{7}%
}\right)  $ to construct the density matrix. Because the concurrence is one
measure of pairwise entanglement, we must trace over the degrees of freedom of
one site in the block. Without loss of generality, we trace over the site 2.
The reduced density matrix for the sites 1 and 3 can be obtained as,%
\begin{equation}
\rho_{13}=\frac{1}{4(\gamma^{2}+1)}\left(
\begin{array}
[c]{cccc}%
2 & 0 & 0 & 2\gamma\\
0 & \gamma^{2}+1 & \gamma^{2}+1 & 0\\
0 & \gamma^{2}+1 & \gamma^{2}+1 & 0\\
2\gamma & 0 & 0 & 2\gamma^{2}%
\end{array}
\right)  . \label{P3}%
\end{equation}
$C_{13}$ denotes the concurrence of the sites 1 and 3 which is defined as
\begin{equation}
C_{13}=\text{Max}\left\{  \lambda_{4}^{1/2}-\lambda_{3}^{1/2}-\lambda
_{2}^{1/2}-\lambda_{1}^{1/2},\ 0\right\}  , \label{con}%
\end{equation}
where the $\lambda_{k}\left(  k=1,2,3,4\right)  $ are the eigenvalues of
$\widehat{R}=\rho_{13}\widetilde{\rho_{13}}\ $[where $\widetilde{\rho_{13}%
}=\left(  \sigma_{1}^{y}\otimes\sigma_{3}^{y}\right)  \rho_{13}^{\ast}\left(
\sigma_{1}^{y}\otimes\sigma_{3}^{y}\right)  $] in ascending order. The
eigenvalues of $\widehat{R}$ can be exactly solved which are%
\begin{equation}
\lambda_{1}=\lambda_{2}=0,\ \lambda_{3}=\frac{\gamma^{2}}{\left(  1+\gamma
^{2}\right)  ^{2}},\ \lambda_{4}=\frac{1}{4}. \label{gg}%
\end{equation}

We substitute the above equations into Eq. (\ref{con}), then $C_{13}$ can be
gotten which is the function of $\gamma.$ The renormalization of $\gamma$
defines the evolution of concurrence on the increasing of the size of the
system. For better to discuss the concurrence, we set $C_{13}$\ as a function
of $g,$ where%
\begin{equation}
g=\left(  1+\gamma\right)  /\left(  1-\gamma\right)  . \label{20}%
\end{equation}
$C_{13}$ versus different QRG iterations is plotted in Fig. 2.

For different QRG steps, the plots of $C_{13}$ versus different QRG steps
cross each other at the critical point. In the thermodynamic limit, the
concurrence develops two saturated values which is nonzero for $g_{c}=1$ and
zero for $0\leq g<1$ and $g>1$. At the point $g_{c}=1$, the system exists
quantum correlation because $C_{13}$ is a nonzero constant. The infinite chain
can be effectively described by a three-site model with the renormalized
coupling constants. In this case, the quantum fluctuations play an important
role and destroy any long-range order of the system. Nonzero $C_{13}$ verifies
that the system is entangled where the ground state is characterized by a
gapless excitation and algebraic decay of spin correlations. While for $0\leq
g<1$ and $g>1$, the model is gap with magnetic long-range order, the
nontrivial points correspond to two Ising phases ordered in $y$ direction
$\left(  g\rightarrow0\right)  $ and $x$ direction $\left(  g\rightarrow
\infty\right)  $, respectively. It can be seen that the N\'{e}el phase is the
dominant phase of the system.

\section{NON-ANALYTIC AND SCALING BEHAVIOR \label{disi}}

The system can occur QPT because of the non-analyticity property of the
block-block entanglement. The non-analyticity behavior can be also accompanied
by a scaling behavior because of the diverging of the correlation length. In
this section, we show the behaviors of the concurrence and the QPT in the XY
model. By using of QRG method, a large system $\left(  N=3^{n+1}\right)  $ can
be effectively described by three-site block with the renormalized coupling
constants after the $n$th iteration of RG. Thus, the entanglement between two
renormalized sites represents the entanglement between two blocks of the
system each containing $N/3$ sites.

The first derivative of concurrence is analyzed and shows a singular behavior
when the size of the system is infinity. All these dates have been shown in
Fig. 3. It is easy to see that the first derivative of the concurrence is
discontinuous at $g_{c}=1$, while $C_{13}$ is continuous. This indicates that
the QPT of the system is the second-order QPT \cite{J.I.2005}. The scaling
behavior of the maximum of $y=dC_{13}/dg$ versus $N$ is plotted in Fig. 4
which shows a linear behavior of ln$\left(  y_{\max}\right)  $ versus
ln$\left(  N\right)  $. In order to show the intuitionistic scaling behavior,
the numerical results are given. The position of maximum $\left(
g_{m}\right)  $ of $y$ touches the critical point as the size of the system
increases. This is plotted in the Fig. 5. It is found that there are the
relation of $g_{m}=g_{c}-N^{-\theta}$ with $\theta$ $=0.98$. The exponent
$\theta$ which is called entanglement exponent is directly related to the
correlation length exponent closing to the critical point. The singular
behavior of the concurrence and the scaling behavior of the system depend on
the entanglement exponent, which is the reciprocal of correlation length
exponent, i.e., $\theta$ $=1/\nu$. The emerging singularity connects to the
universality class of the model. As the critical point is approached in the
limit of large size, the correlation length covers the whole system. That is
to say, the RG implementation of entanglement truly captures the critical
behavior of the XY model in the vicinity of the critical point.

\section{CONCLUSIONS\ \label{diwu}}

In this paper, we discuss the entanglement and QPT in the anisotropic $s=1/2$
XY model by the quantum renormalization group method. The concurrence as one
measure of the quantum correlation is investigated. The critical behavior of
the system is obtained by the renormalization of the lattice. As the number of
RG iterations reaches the infinity, the system occurs QPT between the
spin-fluid and the Ising-like phases which correspond to two different fixed
values of the concurrence at the critical point and both sides of it
respectively. The diverging behavior of the first derivative of the
concurrence is accompanied by the scaling behavior in the vicinity of the
critical point. Further insight, the scaling behavior is investigated and
characterizes how the critical point of the model is touched as the size of
system increases. In the thermodynamic limit, the non-analytic behavior of
entanglement is correlated with the diverging of the correlation length at the
critical point.

\begin{acknowledgments}
This work is supported by the National Natural Science foundation of China
under Grant No. 10775088, the Shandong Natural Science foundation under Grant
No. Y2006A05, and the Science foundation of Qufu Normal University. Fu-Wu Ma
would like to thank Yin-Yang Shen, Sha-Sha Li, and Hong Li for many fruitful
discussions and useful comments.
\end{acknowledgments}

\newpage\ \ \ Figure captions:

Fig. 1. The kadanoff's block renormalization group approach of a chain where
we consider three sites as a block.

Fig. 2. Representation of the evolution of the concurrence versus $g$ in terms
of QRG iterations.

Fig. 3. First derivative of concurrence and its manifestation toward diverging
as the number of QRG iterations increases (Fig. 2).

Fig. 4. The logarithm of the absolute value of maximum, ln$(|dC_{13}%
/dg|g_{\max}|)$, versus the logarithm of chain size, ln$(N)$, which is linear
and shows a scaling behavior. Each point corresponds to the maximum value of a
single plot of Fig. 3.

Fig. 5. The scaling behavior of $g_{\max}$ in terms of system size $\left(
N\right)  $ where $g_{\max}$ is the position of maximum in Fig. 3.

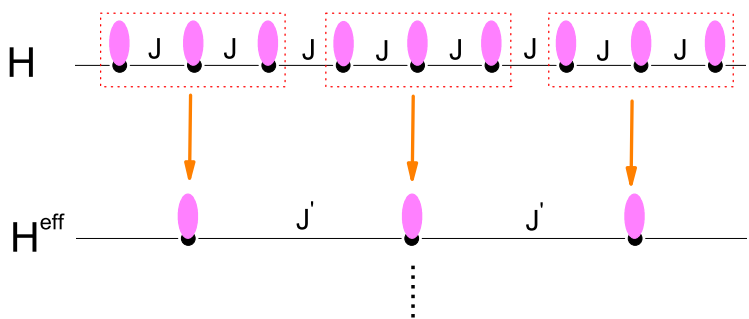
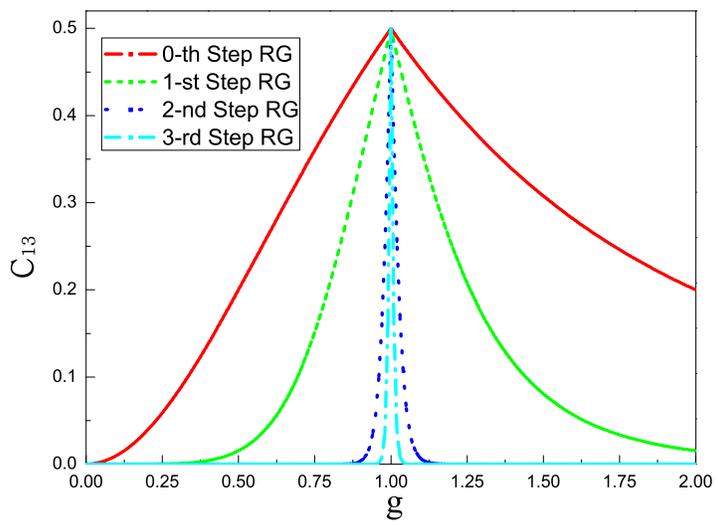
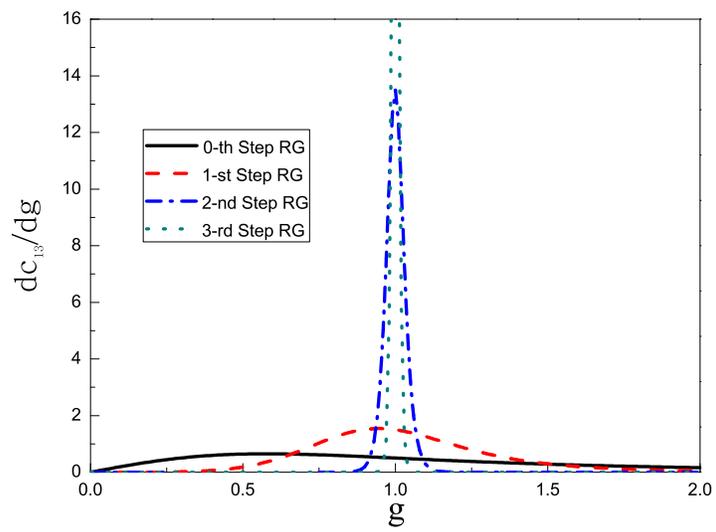
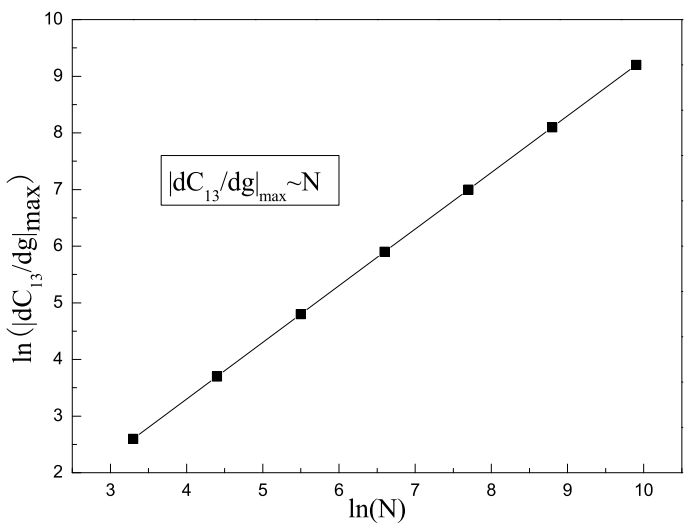
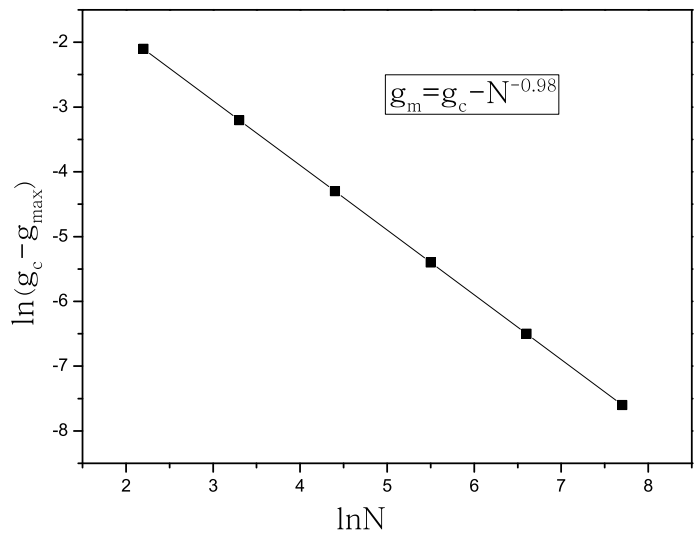
\end{document}